\documentstyle[prd,aps,epsfig]{revtex}
\begin{document}
\draft

\topmargin=-1cm
\twocolumn[\hsize\textwidth\columnwidth\hsize\csname@twocolumnfalse\endcsname
%
\title{Fifteen Minutes of Fame: The Dynamics of Information Access
on the Web}
\author{Z. Dezs\H{o}$^1$, E. Almaas$^1$, A. Luk\'acs$^2$, B. R\'acz$^2$,
I. Szakad\'at$^3$, A.-L. Barab\'asi$^1$}
\address{
{\it 1.Center for Complex Network Research and Department of Physics, University of Notre Dame, Notre Dame,
IN 46556
\newline
2. Computer and Automation Research Institute, Hungarian
Academy of Sciences MTA SZTAKI, Budapest, Hungary
\newline
3.Axelero Internet Provider Inc., 1364 Budapest, Hungary
 }
       }

\maketitle
\centerline{\small (Dated \today)}

\begin{abstract}
While current studies on complex networks focus on systems that
change relatively slowly in time, the structure of the most 
visited regions of
the Web is altered at the timescale from hours to days.
 Here we investigate the
dynamics of visitation of a major news
portal, representing the prototype for such a rapidly evolving network.
The nodes of the network can be classified into 
stable nodes, that form the time independent
skeleton of the portal, and news documents.  
The visitation of the two node classes are markedly different,
the skeleton acquiring visits at a constant rate, while a news document's
visitation peaking after a few hours. We find that the visitation pattern
of a news document decays as
a power law, in contrast with the exponential prediction provided by 
simple models of site visitation.
This is rooted in the inhomogeneous
nature of the browsing pattern characterizing individual users:
the time interval between consecutive visits by the same user
to the site follows a power law distribution, in contrast with the exponential
expected for Poisson processes. We show that the exponent characterizing
the individual user's browsing patterns determines
the power-law
decay in a document's visitation.
 Finally, our
results document the fleeting quality of news and events: while fifteen minutes
of fame is still an exaggeration in the online media, we find that 
access to most news items significantly decays after 36 hours of posting.

\end{abstract}
\vspace{2pc}
]
\vspace{1cm}

The recent interest in
the topological properties of complex networks is driven by
the realization that understanding
the evolutionary processes responsible for network formation
is crucial for comprehending
the topological maps describing many real systems 
\cite{review,mendes_book,vespignani,newman_review,newman_book,tor,oltvai,schuster,strogatz}.
A much studied example is the WWW,
whose topology is driven by its continued expansion
through 
the addition of new documents and links.
This growth process has inspired a series of 
network models
that reproduce some of the most studied topological 
features of the Web \cite{albert_dia,huberman1999,kleinberg,pennock,dorogov,kahng,ba_science,ba_phys}. 
 The bulk of the current topology driven
research focuses on the so called publicly indexable web, which
changes only slowly, and therefore can be reproduced 
with reasonable accuracy.
In contrast, the most visited portion of the WWW, ranging from news
portals to commercial sites, change within hours
through the rapid addition and removal of documents and links.
This is driven by the fleeting quality of news:
 in contrast with the 24-hour news cycle
 of the printed press, in the online media the non-stop 
stream of new developments often obliterates an event within hours.
But the WWW is not the only rapidly evolving network:
the wiring of a cell's regulatory network can also change very 
rapidly during cell cycle or when there are rapid changes in 
environmental and stress factors \cite{oltvai}. 
Similarly, while in social networks
the cumulative number of friends and acquaintances an individual has
is relatively stable, an individual's contact network, representing those 
that it interacts with during a given time interval, 
is often significantly altered 
from one day to the other. Given the widespread occurrence of these rapidly 
changing networks, it is important to understand their topology and dynamical features.

Here we take a first step in this direction by studying as a model system 
a news portal, 
consisting of news items that are added and removed at a rapid rate. 
In particular, we focus on the interplay between the network and the visitation history 
of the individual documents. 
In this context, users are often modeled as random walkers, that
move along the links of the WWW. Most research on diffusion on complex networks  \cite{noh_1,noh_2,jesperson,lahtinen_1,lahtinen_2,pandit,almaas_dif,monasson,huberman1998}
ignores the precise {\it timing} of the 
visit to a particular web document. There are good reasons for this:
such topological quantities as mean free path or probability of return
to the starting point can be expressed using the diffusion 
time, where each time step corresponds to a single diffusion step. Other 
approaches assume that the diffusion pattern is a Poisson process 
\cite{poisson}, so that the probability of an HTML request
in a $dt$ time interval is $pdt$. 
In contrast, 
here we show that the timing of the browsing process is non-Poisson, 
which has a significant impact on the visitation history of web 
documents as well.

\section{Dataset and network structure}
 Automatically assigned cookies allow us to reconstruct the
 browsing history of approximately 250,000 unique visitors
of the largest Hungarian news
and entertainment portal (origo.hu),
which provides online news and magazines, community pages,
software downloads, free email and search engine, capturing 
40\% of all internal Web traffic in Hungary. The portal 
receives 6,500,000 HTML hits on a typical workday.
We used the log files of the portal to collect the visitation pattern of each 
visitor 
between 11/08/02 and 12/08/02,
the number of new news documents released in this time period
being 3,908.

 From a network perspective most web portals consist
 of a stable skeleton,
representing the overall organization of the web portal,
and a large number of news items
that are documents only temporally linked to the skeleton.
Each news item represents a particular web document with
a unique URL.
A typical news item is added to the main page, as well 
as to the specific news subcategories to which it belongs.
For example, the report about an important soccer match could start 
out simultaneously on the front page, the sports page and the 
soccer subdirectory of the sports page.
As a news document ``ages'', 
new developments compete for space, thus the document is gradually removed from 
the main page, then from the sports page and 
eventually from the soccer page as well. 
After some time (which varies from document to document) 
an older news document, 
while still available on the portal, will be disconnected from the skeleton,
and can be
accessed only through a search engine. To fully
understand the dynamics of this network, we need to distinguish between
the stable skeleton and the news documents with heavily time dependent visitation.

The documents belonging to the skeleton are characterized by an
approximately constant daily visitation pattern, thus the 
cumulative number of visitors accessing them increases linearly 
in time.
 In contrast, the visitation of news documents is the highest 
right after their release and decreases in time, thus their 
cumulative visitation reaches a saturation after several days.
This is illustrated in Fig. 1, where we show the cumulative visitation for 
the main page (www.origo.hu/index.html) and a typical news item. 

\begin{figure}[h]
\begin{center}
\epsfig{figure=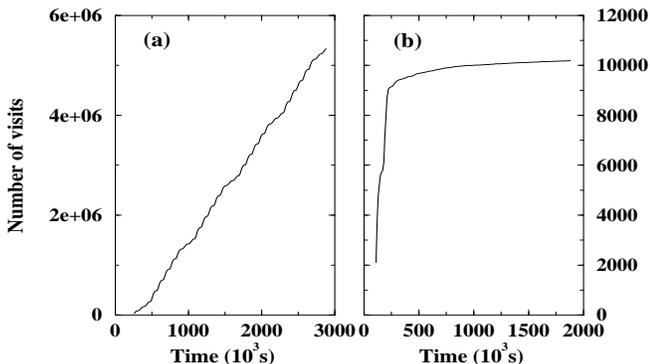,height=5cm,width=8cm}
\caption{The cumulative number of visits to a typical skeleton
document (a) and a news document (b).
The difference between the two visitation patterns
allows us to distinguish between news documents and the stable documents 
belonging to the skeleton.
  }
\label{fig:fig1}
\end{center}
\end{figure}

The difference between the two visitation patterns
allows us to distinguish in an automated fashion
 the websites belonging to the skeleton from the news documents.
For this we 
make a linear regression to each site's cumulative visitation pattern
and calculate the deviation from the fitted lines, 
 documents with very small deviations being assigned to the skeleton.
The validity of the algorithm was checked by inspecting the 
URL of randomly selected documents, as the skeleton and the news documents in most cases
have a different format. 
But given some ambiguities in the naming system, we used the visitation based 
distinction to finalize the classification of the documents into skeleton and news.

When visiting a news portal, we often get the impression that it
has a hierarchical structure.
As shown in Fig. 2  the skeleton forms a complex network,
driving the visitation patterns of the users.
Indeed, the main site, shown in the center, is the most visited, and 
the documents to which it directly links to also represent highly visited sites. In general (with a few notable exceptions, however), the further we go 
from the main site on the network, the smaller is the visitation.  
 The skeleton of the studied portal has 933 documents
with an average degree close to $2$ (i.e. it is largely a tree, with only a few loops, confirming 
our impression of a hierarchical topology), the network
having a few well connected nodes (or hubs), while many are linked 
to the skeleton by a single link \cite{ba_science,ba_phys}.

\begin{figure}[h]
\begin{center}
\epsfig{figure=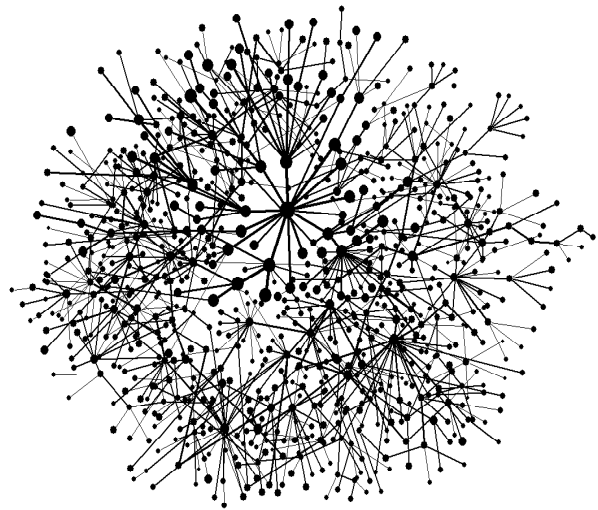,height=7cm,width=9cm}
\caption{The skeleton of the studied web portal
 has 933 nodes.
The area of the circles assigned to each node in the figure is proportional with the logarithm of 
the total number of visits to the corresponding web document. The width 
of the links are proportional with the logarithm of the total
number of times the hyperlink was used by the surfers on the portal.
The central largest node corresponds to the main page 
(www.origo.hu/index.html) directly connected to several other highly 
visited sites. 
}
\label{fig:fig1}
\end{center}
\end{figure}

\section{The dynamics of network visitation}
Given that the difference between the skeleton and the news documents 
is driven by the visitation patterns, next we focus on 
the interplay between 
the visitation pattern of individual users and the overall visitation
of a document.
The overall visitation of a specific document is expected to be determined 
both by the document's position on the web page, as well as the content's 
potential importance for various user groups.
In general the number of visits $n(t)$ 
to a news document follows a dampened periodic pattern: the majority of visits
 (28\%) take place within the first day, decaying to only 7\% on the
 second day, and reaching a small but apparently constant visitation
 beyond four days (Fig 3a).
 Given that after a day or two most news are archived, the
 long-term saturation of 
visitation corresponds to direct search or traffic from
 outside links. 

\begin{figure}[h]
\begin{center}
\epsfig{figure=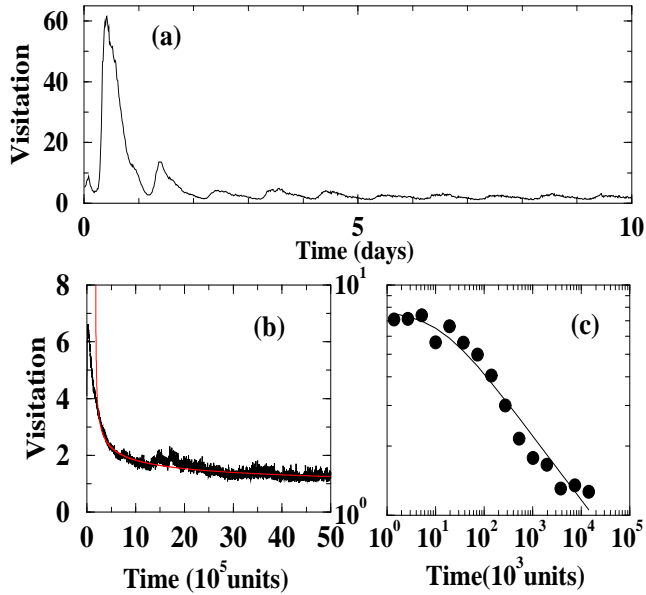,height=8cm,width=8.5cm}
\caption{(a) The visitation pattern of news documents on a web portal.
The data represents an average over 3,908 news
documents, the release time of each being shifted to day one, 
keeping the release hour unchanged. The first peak indicates that most visits 
take place on the release day, rapidly decaying afterward.  
(b) The same as plot (a), but to reduce the daily fluctuations 
we define the time unit as one web page request on the portal.  
(c) Logarithmic binned decay of visitation of (b) shown in a log-log plot, 
indicating that the visitation follows 
$n(t) \sim (t+t_0)^{-\beta}$, with $t_0=12$ and 
$\beta=0.3\pm0.1$ shown as a continuous line on both (b) and (c).
  }
\label{fig:fig1}
\end{center}
\end{figure}

To understand the origin of the observed decay in visitation,
we assume that the portal has $N$ users,
each reading 
the news document of direct interest for him/her.
Therefore,
at every time step each user reads a given document
with probability $p$. Users will not read the same news
more than once, therefore the number of users which have not read a given document decreases with time. 
We can calculate the time dependence of the number of potential readers
to a news document using
\begin{equation}
\frac{d {\cal N} (t)}{dt}=-{\cal N}(t)p
\end{equation}  
where ${\cal N}(t)$ is the number of visitors which have not read the
selected news document by time 
$t$.   
The probability that a new user reads the news document
is given by
${\cal N}(t)p$.
Equation (1) predicts that
\begin{equation}
{\cal N}(t)=N\exp(-t/t_{1/2})
\end{equation}
where $t_{1/2}=1/p$, characterizing the halftime of the news item.
The number of visits ($n$) in unit time is given by

\begin{equation}
n(t)=-\frac{d{\cal N}}{dt}=\frac{N}{t_{1/2}}\exp(-t/t_{1/2}).
\end{equation}

 Our measurements indicate, however, that in contrast with this
 exponential prediction the visitation does not decay 
exponentially, but its asymptotic behavior is best 
approximated by a power law (Fig 3c)
\begin{equation}
n(t) \sim t^{-\beta}
\end{equation} 
with $\beta=0.3 \pm 0.1$, so that while the bulk of the
 visits takes place at small $t$, a considerable number of visits are recorded
 well beyond the document's release time.

 Next we show that the failure of the exponential model is rooted in the uneven browsing
 patterns of the individual users. 
Indeed, Eqs. (1) and (2) are valid only if the users visit the site in regular fashion 
such that they all notice almost instantaneously
a newly added news document.
 In contrast, we find that the time interval
 between consecutive HTML requests by the same visitor is not uniform,
 but follows a power law distribution,
$P(\tau) \sim \tau^{-\alpha}$, with 
$\alpha=1.2 \pm 0.1$ (Fig 4a).
\begin{figure}[h]
\begin{center}
\epsfig{figure=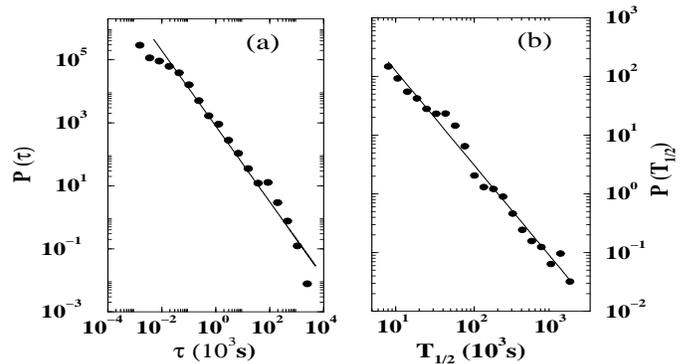,height=5cm,width=8cm}
\caption{(a) The distribution of time intervals between two consecutive 
visits of a given user. The 
cutoff for high $\tau$ ($\tau \approx 10^6$)
captures finite size effects, as time delays over 
a week are undercounted in the month long dataset.
The continuous line has slope $\alpha=1.2$ 
(b) The halftime distribution 
for individual news items, following a power-law with exponent $-1.5\pm 0.1$. }
\label{fig:fig1}
\end{center}
\end{figure}
This means that for each user numerous frequent downloads are
 followed by long periods of inactivity, a bursting, non-Poisson
 activity pattern that is a generic feature of human behavior \cite{barabasi_cikk}
and it is
observed in
 many natural and human driven
dynamical processes \cite{omori1895,abe,alexei,Instant,supercomputers,ftp,economic1,economic2,games,Maya,phone-design,ivanov}.
In the following we show
that this uneven user visitation pattern is responsible for the 
slow decay in the visitation of a 
news document and that $n(t)$ can be derived from the browsing pattern of the 
individual users.

Let us assume that a given news document was released at time $t_0$ and that 
all users visiting the main page after the release read that news. 
Because each user reads each document only once, the visitation of a given document
is determined by the number of {\it new} users visiting the 
page where the document is featured.
\renewcommand{\thefigure}{\arabic{figure}SM}
\renewcommand{\thefigure}{\arabic{figure}}
\begin{figure}[h]
\begin{center}                                                                        

\epsfig{figure=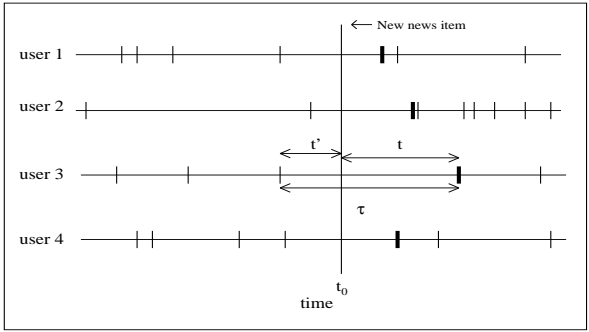,height=4.5cm,width=8cm}
\caption{The browsing pattern of four users, every vertical line representing
the time of a visit to the main page. The time 
a news document was released on the main page
is shown at $t_0$. The thick vertical bars represent
the first time the users visit the main page after the news document was 
released, i.e. the time they could first visit and read the article.}
\label{fig:fig1}
\end{center}
\end{figure}

In Fig. 5 we show the browsing pattern for four different users,
each vertical line representing 
a separate visit to the main page. The thick lines show for each user 
the first time they visit the main page {\it after} the studied 
news document was released at 
$t_0$. 
The release time of the news ($t_0$) divides the time interval $\tau$ 
into two consecutive visits 
of length $t'$ and $t$, where $t+t'=\tau$. 
The probability that a user visits at time $t$
 after the news was released is proportional to the number of possible 
 $\tau$ intervals, which for a given $t$ is proportional to 
 the possible values of $t'$ given by 
 the number of intervals having a length larger than $t$,

\begin{equation}
P(\tau>t)=\int_{t}^{\infty} {\tau}^{-\alpha}d{\tau} \sim t^{-\alpha+1}.
\end{equation}

If we have $N$ users, each following a similar browsing
statistics, the number of new users visiting the main page and reading 
the news item in a unit time ($n(t)$) follows

\begin{equation}
n(t) \sim NP(\tau>t) \sim Nt^{-\alpha+1}.
\end{equation}

Equation (6) connects the exponent $\alpha$ characterizing the decay in the news visitation
 to $\beta$ in Eq. (4), characterizing the visitation pattern 
of individual users, 
providing the relation
\begin{equation}
 \beta=\alpha-1.
\end{equation}
This is in agreement with our measurements within the error bars, as we find
that $\alpha=1.2 \pm 0.1$
and $\beta=0.3 \pm 0.1$.

To further test the validity of our predictions
we studied the relationship between $\alpha$ and $\beta$
for the more general case, when a user that visits the main page reads 
a news item with probability $p$.
We numerically generated browsing patterns for 10,000 users, the distribution
for the time intervals between two consecutive visits, $P(\tau)$, following a power-law 
with exponent $\alpha=1.5$ (Fig. 6 inset).

\renewcommand{\thefigure}{\arabic{figure}}

\begin{figure}[h]
\begin{center}
\epsfig{figure=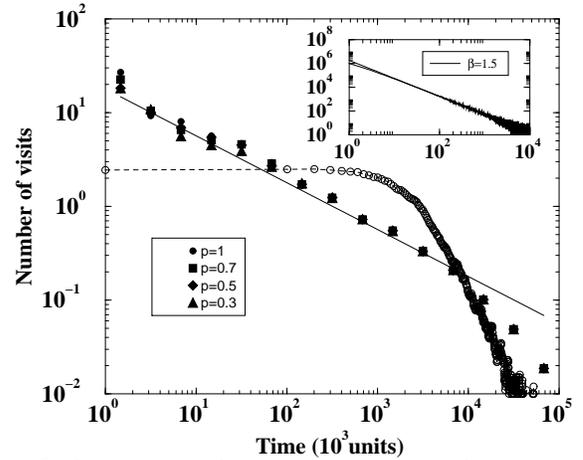,height=6cm,width=7cm}
\caption{ We numerically generated browsing patterns for 10,000 users,
the distribution of the time intervals between two consecutive visits 
by the same user following a power-law with exponent $\alpha=1.5$.
We assume that users visiting the main page will read a given
news item with probability $p$. The number of visits per unit time 
decays as a power-law with exponent $\beta=0.5$ for
four different values of $p$ (circles for $p=1$, squares for
$p=0.7$, diamonds for $p=0.5$ and triangle for $p=0.3$).
The empty circles represent the visitation of a news item if the users 
follow a Poisson browsing pattern. We keep the average time 
between two consecutive visit of each user the same 
as the one observed in the real data.
As the figures indicates, the Poisson browsing pattern cannot reproduce the real visitation decay of a 
document, predicting a much faster (exponential) decay.  
}
\label{fig:fig1}
\end{center}
\end{figure}

In Fig. 6 we calculate the
visits for a given news item, assuming that the users visiting the main page 
read the news with probability $p$, characterizing the
"stickiness" or the potential interest in a news item.
As we see in the figure the value  
of $\beta$ is close to $0.5$ as predicted by (7).
Furthermore, we find that $\beta$ is independent 
of $p$,
indicating that the
inter-event time distribution $P(\tau)$  characterizing 
the individual browsing patterns is the main factor that determines the
visitation decay of a news document,
the difference in the content (stickiness)
of the news playing no significant role. As a reference, we also determined
the decay in the visitation assuming that the users follow a Poisson 
visitation pattern \cite{poisson} with the same inter-event time as 
observed in the real data. As Fig. 6 shows, a Poisson visitation pattern 
leads to a much faster decay in document visitation then the power-law
 seen in Fig. 3c. Indeed, using Poisson inter-event time distribution in
(5) would predict an exponentially decaying tail for $n(t)$.

It is useful to characterize the interest in a news document by its
half time ($T_{1/2}$), corresponding to the time frame during which
half of all visitors that eventually access it have visited. We find
that the overall half-time distribution follows a power law (Fig. 4b),
indicating that while most news have a very short lifetime, a few
continue to be accessed well beyond their initial release. The average
halftime of a news document is 36 hours, i.e. after a day and a half the
interest in most news fades. A similar broad distribution is
observed when we inspect the total number of visits a news document
receives (Fig. 7), indicating that the vast majority of news generate little
interest, while a few are highly popular \cite{menczer2004}. Similar weight
distributions are observed in a wide range of complex networks \cite{goh,vesp,szabo,almaas,yook}. 

\begin{figure}[h]
\begin{center}
\epsfig{figure=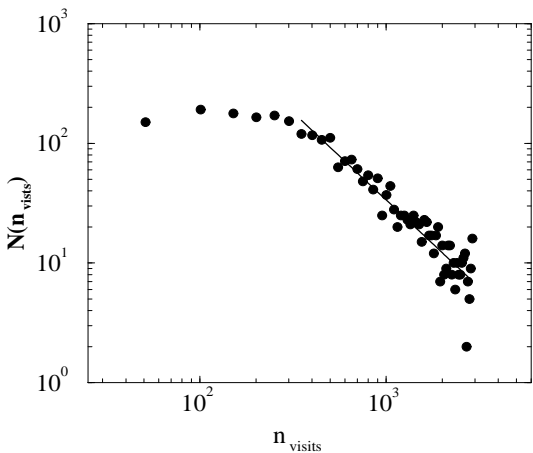,height=6cm,width=7cm}
\caption{ The distribution of the total number of visits different news
documents receive during a month.
 The tail of the distribution follows a power law with 
exponent 1.5.}
\label{fig:fig1}
\end{center}
\end{figure}

The short display time of a given news document,
combined with the uneven visitation pattern indicates that users could miss a significant fraction of the news by not visiting the portal when a document is displayed. We find that a typical user sees only 53\% of all news items appearing on the main page of the portal, and downloads (reads) only 7\% of them. Such shallow news penetration is likely common in all media, but hard to quantify in the absence of tools to track the reading patterns of individuals.

\section{Discussion}
Our main goal in this paper was to explore the interplay between
individual human visitation patterns and the visitation of specific websites
on a web portal. While we often tend to think that the visitation of a given 
document is driven only by its popularity, our results offer a more complex picture:
the dynamics of its accessibility is equally important.
Indeed, while ``fifteen minutes of fame'' does not yet apply to
 the online world, our measurements indicate
 that the visitation of most news items decays
 significantly after 36 hours of posting. The average lifetime must
 vary for different media, but the decay laws we identified are likely
 generic, as they do not depend on content, but are determined mainly by the
 users' visitation and browsing patterns
 \cite{barabasi_cikk}.
These findings also offer a potential explanation of the observation 
that the visitation of a website decreases as a power law following a peak 
of visitation after the site was featured in the media \cite{johansen2000}.
Indeed, the observed power law decay most likely characterizes 
the dynamics of the {\it original} news article, which, 
due to the uneven visitation patterns of the users, displays a power law visitation 
decay (see eq. (4)).
 
These results are likely not limited to news portals.
Indeed, we are
faced with equally dynamic network when we look at commercial sites,
where items are being taken off the website as they are either sold or
not carried any longer.
It is very likely that the visitation of the individual users to such commercial 
sites also follows a power law interevent time, potentially leading to a power law
decay in an item's visitation.
 The results might be applicable to biological
systems as well, where the stable network represents the
skeleton of the regulatory or the metabolic network, indicating which nodes {\it could} interact \cite{almaas,oltvai}, while the
rapidly changing nodes correspond to the actual molecules that are
present in a given moment in the cell. As soon as a molecule is consumed
by a reaction or transported out of the cell, it disappears from the
system. Before that happens, however, it can take place in multiple
interactions.
Indeed, there is increasing experimental evidence that network usage in 
biological systems is highly time dependent \cite{vidal2004,gernstein2004}. 

While most research on information access focuses on search engines
\cite{lawrence1999}, a significant fraction of new information we are exposed to comes from news, whose source is increasingly shifting online from the traditional printed and audiovisual media. News, however, have a fleeting quality: in contrast with the 24-hour news cycle of the printed press, in the online and audiovisual media the non-stop stream of new developments often obliterates a news event within hours. Through archives the Internet offers better long-term search-based access to old events then any other media before. Yet, if we are not exposed to a news item while prominently featured, it is unlikely that we will know what to search for. The accelerating news cycle raises several important questions: How long is a piece of news accessible without targeted search? What is the dynamics of news accessibility? The results presented above
show that the online media allows us to address these questions in a quantitative manner, offering surprising insights into the universal aspects of information dynamics. 
Such quantitative
 approaches to online media not only offer a better understanding of
 information access, but could have important commercial applications
as well,
 from better portal design to understanding information diffusion
\cite{pastor-satorras2001,ciliberti2000,havlin2002}, flow \cite{toroczkai2004} and marketing in the online world.

\end{document}